\documentclass[twocolumn,showpacs,preprintnumbers,amsmath,amssymb]{revtex4}
\usepackage{graphicx}
\usepackage{bm}

\begin{document}

\title{Gravitational self-force correction to the innermost
stable circular orbit of a Schwarzschild black hole}

\author{Leor Barack$^1$ and Norichika Sago$^{1,2}$}
\affiliation
{${}^1$School of Mathematics, University of Southampton, Southampton,
SO17 1BJ, United Kingdom\\
${}^2$Yukawa Institute for Theoretical Physics, Kyoto University,
Kyoto 606-8502, Japan}

\date{\today}

\begin{abstract}
The innermost stable circular orbit (ISCO) of a test particle around a
Schwarzschild black hole of mass $M$ has (areal) radius $r_{\rm isco}=
6M G/c^2$. If the particle is endowed with mass
$\mu(\ll M)$, it experiences a gravitational self-force whose conservative
piece alters the location of the ISCO. Here we calculate the resulting shifts
$\Delta r_{\rm isco}$ and $\Delta\Omega_{\rm isco}$ in the ISCO's radius and
frequency, at leading order in the mass ratio $\mu/M$.
We obtain, in the Lorenz gauge, $\Delta r_{\rm isco}=-3.269 (\pm 0.003)\mu G/c^2$
and $\Delta\Omega_{\rm isco}/\Omega_{\rm isco}=0.4870 (\pm 0.0006) \mu/M$.
We discuss the implications of our result within the context of the
extreme-mass-ratio binary inspiral problem.
\end{abstract}

\maketitle

The last few years have seen a breakthrough in the development
of computational tools for tackling the two-body problem
in general relativity \cite{Pretorius:2005gq}. Numerical-relativistic
(NR) codes can now track the complicated nonlinear dynamics of a spacetime
describing the inspiral and merger of two black holes with comparable masses.
A subset of problems which remains so far beyond the reach of
full NR treatment concerns the radiative
inspiral of a small compact object into a black hole of a much larger mass.
(The study of such {\it extreme-mass-ratio inspirals} is motivated strongly
by the prospects of observing gravitational waves from captures of
compact stars by massive black holes in galactic nuclei---a particularly
exciting source for the planned space-based gravitational wave detector
LISA \cite{Barack:2003fp}.) The large span of lengthscales and long
inspiral timescale render this problem intractable using current NR
technology \cite{Gonzalez:2008bi}.

The problem, however, falls naturally
within the remit of black-hole perturbation theory, which utilizes
the small mass ratio $\eta$ as an expansion parameter. In this approach,
the small object is treated as a structureless point particle, which
moves on the fixed background of the massive hole. At the test-mass
limit ($\eta\to 0$) the particle traces a geodesic of the black hole geometry.
When $\eta$ is finite (yet small), the particle interacts with
its own gravitational field, giving rise
to a gravitational self force (SF). The SF has a dissipative piece which
drives the radiative inspiral; it also has a conservative component which
affects the evolution of the orbital phases.

How to regularize the gravitational self-interaction in curved spacetime
is now a well understood problem in fundamental relativity
\cite{Mino:1996nk}, and a program to calculate the gravitational
SF in the strong-field exterior of a Kerr black hole is well underway.
The first actual quantitative results were obtained for radial plunge
\cite{Barack:2002ku} and circular orbits \cite{Barack:2007tm} in a Schwarzschild
exterior. In the circular orbit case, accurate computations of
conservative SF effects were performed using two independent methods,
successfully tested against each other and against results from
post-Newtonian (PN) theory \cite{Detweiler:2008ft,Sago:2008id}.
The authors of this letter have recently
completed the development of a computational platform for calculations
of the SF for generic (bound, eccentric) orbits in Schwarzschild.
Details of this work, which we consider a significant milestone in the
SF program, are given in a forthcoming paper \cite{prep}.

As a first application of our code, we report here the values of the
$O(\eta)$ shifts in the location and frequency of the innermost
stable circular orbit (ISCO) of a Schwarzschild black hole due to
the conservative piece of the gravitational SF. This represents a first
concrete calculation of a conservative SF effect with a clear physical
interpretation and (potentially) observable consequences. The value of
the ISCO frequency, which is gauge-invariant (within a class of
physically-reasonable gauges---see below), has been utilized in the past
as a convenient reference for comparison and calibration
of PN schemes \cite{Damour:2000we}.
Our result is exact at $O(\eta)$ (within a controlled numerical error),
and we expect it to provide an accurate benchmark against which
future PN calculations (and potentially also future NR results for
binaries with small $\eta$) could be tested.

The radiative decay of a physical circular orbit proceeds through an
``adiabatic'' inspiral epoch, during which gravitational radiation reaction
slowly removes energy and angular momentum from the system, on to a brief
plunge episode in which the particle drops into the black hole
along a nearly-geodesic trajectory. The transition from inspiral to plunge
occurs at radii localized around the ISCO. The properties of the transition
regime were studied by Ori and Thorne (OT) \cite{Ori:2000zn} (for circular
equatorial orbits in Kerr), with later generalizations by O'Shaughnessy
\cite{O'Shaughnessy:2002ez} and Sundararajan \cite{Sundararajan:2008bw}.
OT showed, in particular, that the frequency bandwidth of the radiative
transition scales as $\sim\eta^{2/5}$: The transition becomes more
``abrupt''---less gradual---with smaller $\eta$. OT (as also
\cite{O'Shaughnessy:2002ez,Sundararajan:2008bw}) did not tackle the local
SF directly. Instead, they relied on (numerical) computations of the
flux of energy in the radiated gravitational waves, from which the
dissipative piece of the SF was inferred indirectly.
OT had to ignore the conservative piece of the SF, yet unknown
at the time of their analysis. Our work is complementary to OT's
in that we consider {\em only} conservative effects and ignore dissipation:
it is only then that the ISCO becomes {\em precisely} localizable.
As we shall conclude below, our result justifies a posteriori the
omission of conservative SF terms in the OT analysis (at least in the
special Schwarzschild case).

In what follows we (i) derive formulae for the conservative $O(\eta)$
shift in the ISCO radius and frequency in terms of the conservative
piece of the SF; (ii) describe in brief the numerical method used for
computing the conservative SF (delegating details to a forthcoming
paper \cite{prep}); and (iii) state our results and discuss their implications.
In the rest of this letter we use standard geometrized units, with $G=c=1$.
We denote by $M$
the mass of the background Schwarzschild black hole and by $\mu$ the mass
of the particle (hence $\mu/M=\eta$). The particle's orbit is described in
standard Schwarzschild coordinates by
$t=t_{\rm p}(\tau)$, $r=r_{\rm p}(\tau)$ and $\varphi=\varphi_{\rm p}(\tau)$,
where $\tau$ is proper time; without loss of generality we take
$\theta_{\rm p}=\pi/2$.
The metric signature is ${-}{+}{+}{+}$,
and tensorial indices are raised and lowered using the background metric.

{\it Formula for the ISCO shift:}---
At the limit $\mu\to 0$ the particle traces a timelike
geodesic of the Schwarzschild geometry, described
(using overhats to distinguish the geodesic from the
perturbed orbit discussed later) by
\begin{equation} \label{eq10}
\dot{\hat r}_{\rm p}^2=E^2-V_{\rm eff}(\hat r_{\rm p},L^2),
\end{equation}
\begin{equation} \label{eq20}
\dot{\hat t}_{\rm p}= E/f(\hat r_{\rm p}), \quad\quad
\dot{\hat \varphi}_{\rm p}=L/\hat r_{\rm p}^2.
\end{equation}
Here an overdot denotes $d/d\hat\tau$, $E$ and $L$ are, respectively,
the specific energy and angular momentum (constants of the motion),
$f(r)=1-2M/r$, and the radial effective potential is
$V_{\rm eff}(r,L^2)=f(1+L^2/r^2)$.
Bound orbits exist only for $L>\sqrt{12}\,M$, in which case $V_{\rm eff}$
has two extremum points and $E^2-V_{\rm eff}$ can have 3 distinct roots.
In the latter case, the radial motion is bounded as $r_{\rm min}\leq
\hat r_{\rm p}(\hat\tau)\leq r_{\rm max}$,
where $r_{\rm min}$ (``periapsis'') and $r_{\rm max}$ (``apoapsis'') are the
second-largest and largest roots, respectively. For such orbits we define
the ``eccentricity'' $e$ and ``semi-latus rectum'' $r_0$ through
\begin{equation} \label{eq30}
r_{\rm min}=r_0/(1+e), \quad\quad r_{\rm max}=r_0/(1-e).
\end{equation}
The orbits are fully parameterized (up to initial conditions) by any of the
pairs $\{E,L\}$, $\{r_{\rm max},r_{\rm min}\}$, or $\{e,r_0\}$.
{\it Stable circular orbits} occur when $E^2$ equals the minimum of
$V_{\rm eff}$; they have $e=0$ and radius
$r_{\rm min}=r_{\rm max}=r_0=(L^2+L\sqrt{L^2-12M^2})/(2M)$.
No bound orbits exist for $L<\sqrt{12}\,M$.
The circular orbit with $L=\sqrt{12}\,M$ is the ISCO;
it has radius $r_0=6M$.

The nature of the ISCO as a separatrix for stable circular orbits
is understood from the following dynamical consideration. Still working
at $\mu\to 0$, consider a small-$e$ perturbation of a circular geodesic
with radius $r_0$, such that the resulting slightly-eccentric geodesic
has turning points $r_{\rm min}$ and $r_{\rm max}$ as in Eq.\
(\ref{eq30}). For this orbit we write
\begin{equation} \label{eq40}
\hat r_{\rm p}(\hat\tau)=r_0+e\, \hat r_1(\hat\tau)+O(e^2),
\end{equation}
where $\hat r_1(\hat\tau)$ is to be determined below. The $r$ component of the
test-particle's geodesic equation reads
\begin{equation} \label{eq50}
\ddot{\hat r}_{\rm p}=-\frac{1}{2}\,\frac{\partial V_{\rm eff}}
{\partial \hat r_{\rm p}}\equiv{\cal F}_{\rm eff}(\hat r_{\rm p},L^2),
\end{equation}
${\cal F}_{\rm eff}$ being an effective radial force (per unit test mass)
for the geodesic motion. The linear variation of ${\cal F}_{\rm eff}$ with
respect to $e$ (with fixed $r_0$ and $\hat\tau$)
is $\delta_e {\cal F}_{\rm eff}=
(\partial {\cal F}_{\rm eff}/\partial \hat r_{\rm p}) \delta_e \hat r_{\rm p}
+(\partial {\cal F}_{\rm eff}/\partial L) \delta_e L$,
where the partial derivatives are evaluated at $e=0$, and where
$\delta_e \hat r_{\rm p}=e \hat r_1$. $\delta_e L$ is determined by solving
$E^2=V_{\rm eff}(r_{\rm \min},L^2)$ and $E^2=V_{\rm eff}(r_{\rm \max},L^2)$
simultaneously for $E$ and $L$, and then
using Eqs.\ (\ref{eq30}); this gives $L^2=Mr_0^2/(r_0-3M-Me^2)$
and hence $\delta_e L=0$. At $O(e)$, therefore, Eq.\ (\ref{eq50})
reads
\begin{equation} \label{eq80}
\ddot{\hat r}_1(\hat\tau)=-\hat\omega_r^2\, \hat r_1(\hat\tau),
\end{equation}
with
\begin{equation} \label{eq90}
\hat\omega_r^2=-\left.\frac{\partial{\cal F}_{\rm eff}}{\partial \hat r_{\rm p}}
\right|_{\hat r_{\rm p}=r_0}=\frac{M(r_0-6M)}{r_0^3(r_0-3M)}.
\end{equation}
Thus, the $O(e)$ radial motion is simple-harmonic with frequency
$\hat\omega_r$. The orbit is stable under small-$e$ perturbations when
$\hat\omega_r^2>0$, namely for $r_0>6M$ (no circular timelike geodesics exist
for $r_0\leq 3M$). The condition $\hat\omega_r=0$ identifies the ISCO at
$r_0=6M$.

We now turn to consider the $O(\mu)$ conservative correction to
the orbit. The physical SF along the geodesic $\hat r_{\rm p}(\hat\tau)$
is constructed (as prescribed in \cite{Mino:1996nk}) from the retarded
metric perturbation associated with the particle. We denote
this SF by $\mu F_{\rm ret}^\alpha$, and further denote by
$\mu F_{\rm adv}^\alpha$ the force derived in just the same way from
the advanced perturbation [we factor out $\mu$ for later
convenience, noting our $F$'s($\propto\mu$) describe self
acceleration]. The {\it conservative} piece of the SF (per unit $\mu$) is
defined through $F^\alpha\equiv (F_{\rm ret}^\alpha+F_{\rm adv}^\alpha)/2$.
It is convenient to take $\hat\tau=0$ at a periapsis of $\hat r_{\rm p}(\hat\tau)$,
in which case we have the symmetry relation \cite{sym}
$F_{\rm adv}^\alpha(\hat\tau)=\epsilon_{\alpha}F_{\rm ret}^\alpha(-\hat\tau)$
(no summation over $\alpha$), where $\epsilon_{\alpha}=(-1,1,1,-1)$ in
Schwarzschild coordinates. This gives us a practical formula for
extracting the conservative piece from the full (retarded) SF:
\begin{equation} \label{eq112}
F^{\alpha}(\hat\tau)=
\left[F_{\rm ret}^\alpha (\hat\tau)
+\epsilon_{\alpha}F_{\rm ret}^\alpha (-\hat\tau)\right]/2.
\end{equation}
Notice that, since both $F^r$ and $\hat r_{\rm p}$ are periodic and 
even in $\hat\tau$ [and $\hat r_{\rm p}(\hat\tau)$ is monotonic on 
$0\leq\hat\tau\leq\pi/\hat\omega_r$], we may express $F^r$ as a function 
of $\hat r_{\rm p}$ only (for given $r_0,e$). We write 
$F^r=F^r(\hat r_{\rm p};r_0,e)$.

Under the effect of $F^{\alpha}$, the nontrivial components of
the equation of motion read
\begin{equation} \label{eq100}
\ddot r_{\rm p}={\cal F}_{\rm eff}(r_{\rm p},L^2)+ F^r,
\end{equation}
\begin{equation} \label{eq110}
\dot E=- F_t, \quad\quad
\dot L= F_\varphi,
\end{equation}
where $E\equiv \dot{t}_{\rm p}/f(r_{\rm p})$ and
$L\equiv \dot{\varphi}_{\rm p}/r^2_{\rm p}$ are (in general) no 
longer constant along the orbit, and hereafter an overdot denotes 
$d/d\tau$. From symmetry, the perturbed orbit remains equatorial. 
We assume $r_{\rm p}(\tau)$ remains bounded between some $r_{\min}$ 
and $r_{\max}$, define $r_0$ and $e$ as in Eq.\ (\ref{eq30}), and
take $\tau=0$ at a periapsis. Then we may write, through $O(\mu)$, 
$F^r=F^r(r_{\rm p};r_0,e)$. By virtue of Eqs.\ (\ref{eq112}) and
(\ref{eq110}) we similarly have $E=E(r_{\rm p};r_0,e)$ and 
$L=L(r_{\rm p};r_0,e)$ through $O(\mu)$.


Consider first {\em circular orbit} solutions of the set
(\ref{eq100},\ref{eq110}). In this case $E$ and $L$
are constant along the orbit. Their values are obtained by solving
Eqs.\ (\ref{eq10}) (with $\hat r_{\rm p}\to r_{\rm p}$) and (\ref{eq100})
simultaneously, with $\dot r_{\rm p}=\ddot r_{\rm p}=0$ and $r_{\rm p}=r_0$.
This gives
\begin{eqnarray} \label{eq115}
E_0^2 &=&r_0f_0^2(r_0-3M)^{-1}\left[1-(r_0/f_0)F_0^r\right],
\\ \label{eq120}
L_0^2 &=&Mr_0^2(r_0-3M)^{-1} \left[1-(r_0^2/M) F_0^r \right],
\end{eqnarray}
where subscripts `0' denote circular-orbit values.

To identify the new location of the ISCO we again consider a
small-$e$ perturbation of the circular orbit. Writing $r_{\rm p}(\tau)$
as in Eq.\ (\ref{eq40}), we find that $r_1(\tau)$ again satisfies
an equation of the form (\ref{eq80}), where the frequency is now
\begin{equation} \label{eq130}
\omega_r^2=-\frac{d}{dr_{\rm p}}
\left[{\cal F}_{\rm eff}(r_{\rm p},L^2(r_{\rm p}))+ F^r(r_{\rm p})\right]
_{r_{\rm p}=r_0}.
\end{equation}
In obtaining this result we have replaced the linear variations
$\delta_e L$ and $\delta_e F^r$ with $r_1\delta_{r_{\rm p}} L$ and
$r_1 \delta_{r_{\rm p}} F^r$, respectively, making use of the fact that,
through $O(e)$, $L$ and $F^r$ depend on $e$ only through $r_{\rm p}(\tau;r_0,e)$.
That this is true is shown as follows. Let us formally expand
$L=L(r_{\rm p};r_0,e)=L^{(0)}(r_{\rm p};r_0)+eL^{(1)}(r_{\rm p};r_0)+O(e^2)$,
where the coefficients $L^{(n)}$ depend on $e$ only through
$r_{\rm p}$, and $O(e^2)$ represents terms whose explicit dependence on
$e$ is at least quadratic. Now note from Eq.\ (\ref{eq30}) that formally
replacing $e\to -e$ is equivalent to replacing $r_{\rm min}\leftrightarrow
r_{\rm max}$; hence
$L(r_{\rm min};r_0,\mp e)=L(r_{\rm max};r_0,\pm e)$, giving
$L^{(1)}(r_{\rm min};r_0)=-L^{(1)}(r_{\rm max};r_0)$.
Assuming $L$ is a continuous function of $r_{\rm p}$, the last result
implies that $L^{(1)}=O(e)$ for all $r_{\rm p}$.
We thus have $L=L^{(0)}(r_{\rm p};r_0)+O(e^2)$, with the conclusion
that, working at $O(e)$, $L$ may depend on $e$ only through $r_{\rm p}$.
The same argument---with the same conclusion---applies to $F^r$.

It is now useful to describe the $O(e)$ orbit more explicitly: Assuming
periapsis at $\tau=0$, Eq.\ (\ref{eq80}) (overhats removed) integrates 
to give $r_1=-r_0\cos\omega_r\tau$, hence
\begin{equation} \label{eq140}
r_{\rm p}(\tau)=r_0(1-e\cos\omega_r\tau) +O(e^2).
\end{equation}
We can then expand $L(r_{\rm p})$ through $O(e)$ in the form
$L=L_0+er_1 L'(r_0)=L_0-er_0 L'(r_0)\cos\omega_r\tau$,
where a prime denotes $d/dr_{\rm p}$ and $L_0$ is the circular-orbit
value, given in Eq.\ (\ref{eq120}). Since $F_\varphi=\dot{L}$,
we obtain the form
\begin{equation} \label{eq150}
F_\varphi=e\omega_r F^1_{\varphi}\sin\omega_r\tau  +O(e^2),
\end{equation}
where $F^1_{\varphi}\equiv r_0 L'(r_0)$. Similarly, we expand
$F^r(r_{\rm p})=F^r_0+er_1  {F^r}'(r_0)
+O(e^2)$, giving
\begin{equation} \label{eq160}
F^r=F^r_0+e F^r_1\cos\omega_r\tau+O(e^2),
\end{equation}
where $F^r_1\equiv -r_0 {F^r}'(r_0)$. Replacing
$ L'(r_0)\to F^1_{\varphi}/r_0$ and ${F^r}'(r_0)\to -F^r_1/r_0$ in Eq.\
(\ref{eq130}) and substituting for $L_0$ from Eq.\ (\ref{eq120}),
we finally obtain
\begin{equation} \label{eq170}
\omega_r^2=\hat\omega_{r}^2+
\alpha(r_0) F^r_0+\beta(r_0) F^r_1+\gamma(r_0) F^1_{\varphi},
\end{equation}
where
$\alpha=-3r_0^{-1}(r_0-4M)/(r_0-3M)$, $\beta=r_0^{-1}$ and
$\gamma=-2r_0^{-4}[M(r_0-3M)]^{1/2}$.
This formula describes the $O(\mu)$ conservative shift in the radial
frequency off its geodesic value. Note it requires knowledge
of the SF through $O(e)$.

The perturbed ISCO radius $r_{\rm isco}$ is now obtained from the
condition $\omega_r^2(r_{\rm isco})=0$. Recalling Eqs.\ (\ref{eq90})
this gives
\begin{eqnarray} \label{eq180}
\Delta r_{\rm isco}&\equiv& r_{\rm isco}-6M \nonumber\\
&=& \left. (r_0^3/M)(3M-r_0)(\alpha F^r_0+
\beta F^r_1+\gamma F^1_{\varphi})\right|_{r_0=6M}\nonumber\\
&=& 216M^2F^r_{0\rm is} -108M^2 F^r_{1\rm is} +
\sqrt{3}\,F^1_{\varphi\rm is},
\end{eqnarray}
where in the second line we were allowed to replace $r_{\rm isco}\to 6M$
since the SF terms are already $O(\mu)$ [so the error introduced affects
$\Delta r_{\rm isco}$ only at $O(\mu^2)$]. We have denoted
$F^r_{0\rm is}\equiv F^r_0(r_0=6M)$ and similarly for
$F^r_{1\rm is}$, $F^1_{\varphi\rm is}$.
Note  $\Delta r_{\rm isco}$
is {\em gauge dependent}, just like the SF itself \cite{Barack:2001ph}.

A more physically-meaningful quantity associated with the circular orbit
is the azimuthal frequency $\Omega \equiv d\varphi_{\rm p}/dt=
\dot\varphi_{\rm p}/\dot t_{\rm p}$. As discussed in \cite{Sago:2008id},
$\Omega$ is invariant under all $O(\mu)$ gauge transformations whose generators
respect the helical symmetry of the circular-orbit configuration \cite{gauge}.
An expression for the SF-corrected $\Omega$ is obtained using Eqs.\
(\ref{eq115}) and (\ref{eq120}):
\begin{equation}  \label{eq190}
\Omega =\hat\Omega\left[1-\frac{r_0(r_0-3M)}{2 Mf_0}
F_0^r\right],
\end{equation}
where $\hat\Omega\equiv (M/r_0^3)^{1/2}$ is the geodesic (no SF)
value. Evaluated at $r_0=r_{\rm isco}$, the SF-induced frequency shift
$\Delta\Omega\equiv\Omega-\hat\Omega$ reads
{[}through $O(\mu)$]
\begin{equation} \label{eq200}
\Delta\Omega_{\rm isco}
= -\Omega_{\rm isco}\left[
\frac{1}{4}\Delta r_{\rm isco}/M
  + \frac{27}{2}MF_{0\rm is}^r
\right],
\end{equation}
where $\Omega_{\rm isco}\equiv \hat\Omega(r_0=6M)=
(6^{3/2}\,M)^{-1}$.
Equations analogous to (\ref{eq180}) and (\ref{eq200}) were obtained by
Diaz-Rivera {\em et al.}\ \cite{DiazRivera:2004ik} in their study of
scalar SF effects.

{\it Numerical calculation of the SF:}---
Our numerical technique for computing the SF is described in detail
in Refs.\ \cite{Barack:2007tm} (circular orbits) and \cite{prep}
(eccentric orbits). The method is based on a direct integration of
the linearized (hyperpoblic) set of Einstein equations in the Lorenz
gauge, with a delta-function source representing the stress-energy of
the moving particle. The set of 10 coupled equations is
decomposed into tensor-harmonic ($l,m$) modes, and we solve for each
$l,m$-mode of the (retarded) metric perturbation using finite differentiation on
a characteristic mesh in 1+1-dimensions (time+radius). The values of the
metric perturbation modes and their derivatives at the particle are then
fed into the ``mode-sum formula'' \cite{Barack:1999wf,Barack:2001gx},
which yields the physical (retarded) SF through a mode-by mode regularization.
The conservative piece $F^{\alpha}$ is finally constructed using
Eq.\ (\ref{eq112}).

For our ISCO analysis we need the values of the SF coefficients
$F^r_{0}$, $F^r_{1}$ and $F^1_{\varphi}$, evaluated at $r_0=6M$.
These are extracted by fitting our numerical SF data along
the orbit using the analytic formulae (\ref{eq150}) and (\ref{eq160}).
We used two independent methods to obtain the necessary data.
The first is based on applying our eccentric-orbit code \cite{prep} for
a sequence of slightly-eccentric geodesics approaching the ISCO. The second
method (following the treatment of Diaz-Rivera {\it et al.}~in the scalar-field
case \cite{DiazRivera:2004ik}) is based on expanding
the perturbation equations about a circular orbit and solving through
$O(e)$ for a sequence of circular geodesics approaching $r_0=6M$.
In both methods, the ISCO limit is quite delicate (as the radial period
tends to infinity there) and requires a careful treatment---see
\cite{prep} for a detailed discussion of our numerical procedure
and error analysis. Reassuringly, our two methods produce
mutually consistent numerical data.

{\it Results and discussion:}---We obtained
$F^r_{0\rm is}=0.0244666\,\mu/M^2$,
$F^r_{1\rm is}=0.06209\,\mu/M^2$ and
$F^1_{\varphi \rm is}=-1.07\,\mu$.
Equations (\ref{eq180}) and (\ref{eq200}) then give, respectively,
\begin{equation} \label{eq300}
\Delta r_{\rm isco}=-3.269\,\mu,
\quad\quad
\frac{\Delta\Omega_{\rm isco}}{\Omega_{\rm isco}}=0.4870\, \mu/M.
\end{equation}
We estimate \cite{prep} the absolute numerical error in these figures
as $\pm 0.003\,\mu$ and $\pm 0.0006\,\mu/M$, respectively. This error
is controllable and can be reduced in the future simply by running 
our code for longer.
Recall the quoted values correspond to the Lorenz gauge.
Note the ISCO frequency {\it increases} under the conservative
effect of the gravitational SF. This is qualitatively similar to
the effect of the {\em scalar} SF acting on a particle endowed with
scalar charge, which also increases $\Omega_{\rm isco}$
\cite{DiazRivera:2004ik}.

It is interesting to compare the conservative shift
$\Delta\Omega_{\rm isco}$ with the frequency bandwidth of the
dissipative transition across the ISCO. From the analysis
of OT \cite{Ori:2000zn} one estimates the latter
(in the Schwarzschild case) as
$\Delta\Omega_{\rm diss}/\Omega_{\rm isco}\cong 4.3874\, \eta^{2/5}$,
where, recall, $\eta=\mu/M$. Hence,
$\Delta\Omega_{\rm diss}/\Delta\Omega_{\rm isco}\cong 9\, \eta^{-3/5}$,
giving, for example, $\sim 35830$, $9000$ and $2261$ for mass ratios $\eta=10^{-6}$,
$10^{-5}$ and $10^{-4}$, respectively. Thus, for $\eta$ in the
astrophysically-relevant range, the conservative piece of the SF shifts
the ISCO frequency by an amount much smaller than the entire bandwidth
of the transition regime. This confirms what one expects based simply
on the above $\eta^{-3/5}$ scaling, and it provides a firm justification
for omitting the conservative SF in the OT analysis.

The main practical value of our result, we believe, is in that it
provides an accurate strong-field benchmark to inform the
development of approximate methods, such as PN. A particularly
promising PN treatment of the small-$\eta$ inspiral problem is based
on the Effective One Body (EOB) approach \cite{Buonanno:1998gg},
with a recent study \cite{Damour:2007xr} convincingly demonstrating
the effectiveness of EOB in describing the dissipative part of the strong-field
dynamics. It would be interesting to test the performance of the
conservative EOB dynamics against the results obtained here.
We also envisage that our results could help inform the
refinement of approximate ``Kludge'' models for gravitational
waveforms from astrophysical inspirals \cite{Babak:2006uv}.

We acknowledge support from PPARC/STFC through grant number PP/D001110/1.
NS also acknowledges support from Monbukagaku-sho Grant-in-Aid
for the global COE program ``The Next Generation of Physics,
Spun from Universality and Emergence''.

\end{document}